\documentclass[conference]{IEEEtran}
\PassOptionsToPackage{table}{xcolor}  

\usepackage[utf8]{inputenc}
\usepackage{cite}
\usepackage{url}
\usepackage{enumitem}
\usepackage{graphicx}
\usepackage{float}
\usepackage{tabularx}
\usepackage{multicol}
\usepackage[caption=false,font=footnotesize]{subfig}
\usepackage[a4paper, margin=0.7in]{geometry} 

\usepackage{booktabs}
\usepackage{siunitx}
\usepackage{tcolorbox}
\usepackage{makecell}
\usepackage{amsmath,amssymb,amsfonts}
\usepackage{algorithmic}
\usepackage{textcomp}
\usepackage{xcolor}              
\usepackage{array}
\usepackage{multirow}
\usepackage{adjustbox}

\usepackage[colorlinks=true,linkcolor=blue,citecolor=blue,urlcolor=blue]{hyperref}

\begin{document}


\title{ForensicsData: a Digital Forensics Dataset for Large Language Models}

\author{\IEEEauthorblockN{Youssef Chakir\IEEEauthorrefmark{1},
Iyad Lahsen-Cherif\IEEEauthorrefmark{1}}
\IEEEauthorblockA{\IEEEauthorrefmark{1}INPT, CS department, Rabat, Morocco. \\ 
Emails: chakir.youssef@master.inpt.ac.ma, lahsencherif@inpt.ac.ma}
}

\maketitle


\begin{abstract}
The growing complexity of cyber incidents presents significant challenges for digital forensic investigators, especially in evidence collection and analysis. Public resources are still limited because of ethical, legal, and privacy concerns, even though realistic datasets are necessary to support research and tool developments. To address this gap, we introduce ForensicsData, an extensive Question-Context-Answer (Q-C-A) dataset sourced from actual malware analysis reports. It consists of more than 5,000 Q-C-A triplets.  A unique workflow was used to create the dataset, which extracts structured data, uses large language models (LLMs) to transform it into Q-C-A format, and then uses a specialized evaluation process to confirm its quality. Among the models evaluated, Gemini 2 Flash demonstrated the best performance in aligning generated content with forensic terminology.  ForensicsData aims to advance digital forensics by enabling reproducible experiments and fostering collaboration within the research community.
\end{abstract}



\section{Introduction}
Digital forensics has emerged as a critical discipline in modern cybersecurity, focusing on the systematic collection, preservation, examination, and analysis of digital evidence to support legal proceedings and incident response. As digital devices become ubiquitous, forensic investigators face mounting challenges in managing vast quantities of heterogeneous data, making traditional manual analysis methods labor-intensive and error-prone. The standardized forensic process involves identifying evidence sources, preserving data integrity through forensic imaging, collecting information with specialized tools, conducting detailed examinations, and producing legally admissible reports.
However, the development and validation of digital forensic tools face a significant bottleneck: the scarcity of realistic, publicly available datasets for training and testing purposes. This limitation stems from stringent privacy regulations, legal restrictions on data sharing, and the inherently sensitive nature of forensic evidence. Consequently, researchers struggle to access sufficient training data, hampering the development of robust forensic tools and limiting research reproducibility. This challenge is particularly acute in malware analysis, where dynamic threats and evolving attack techniques demand continuous updates to detection capabilities.

The emergence of Large Language Models (LLMs) represents a paradigm shift in artificial intelligence, offering unprecedented capabilities in natural language understanding and generation. Built on transformer architectures and trained on extensive text corpora, state-of-the-art models such as GPT-4, Claude, LLaMA, and Gemini demonstrate remarkable proficiency in complex tasks including text synthesis, summarization, anomaly detection, and information retrieval. In digital forensics, LLMs show promising applications in automating evidence triage, streamlining malware analysis workflows, generating comprehensive investigative reports, and detecting anomalous patterns in digital artifacts.

A particularly compelling advantage of LLMs lies in their ability to generate realistic synthetic datasets that preserve the linguistic and structural properties of authentic forensic data. This capability addresses the critical challenge of dataset scarcity by enabling the creation of training and testing resources without relying on sensitive real-world evidence. Synthetic datasets can maintain the complexity and diversity necessary for robust tool development while circumventing ethical, legal, and privacy constraints.
Despite these promising developments, fundamental questions remain unanswered regarding the effective application of LLMs in digital forensics dataset generation. The quality, diversity, and accuracy of synthetic datasets must be rigorously evaluated to ensure their utility for training forensic tools.
This research addresses these challenges through a comprehensive investigation of LLM-based synthetic dataset generation for digital forensics, with particular emphasis on malware behavior analysis.
Specifically, we evaluate and validate the quality, realism, and effectiveness of our newly introduced \textbf{ForensicsData} dataset.
We focus on three primary research questions:
\begin{enumerate}
    \item \textbf{Dataset Quality and Realism}: Can LLMs generate realistic, diverse, and accurate synthetic malware behavior datasets that effectively capture the complexity and variability of real-world threats?
    \item \textbf{Comparative Model Performance}: How do different LLM architectures compare in terms of generation quality, accuracy, efficiency, and cost-effectiveness for forensic dataset creation?
    \item \textbf{Validation Methodologies}: What validation techniques are most effective for ensuring the quality, reliability, and forensic relevance of synthetic datasets?

\end{enumerate}
 
To address these questions, we present \textbf{ForensicsData}, a comprehensive digital forensics Question-Context-Answer (Q-C-A) dataset derived from contemporary malware analysis reports. This dataset represents the first publicly available, structured Q-C-A resource specifically designed for digital forensics applications, comprising over 5,000 annotated triplets extracted from malware reports published in 2025. Each entry encompasses critical forensic information including malware metadata, behavioral patterns, indicators of compromise (IOCs), tactics, techniques, and procedures (TTPs), and mitigation strategies.

Our research makes three significant contributions to the digital forensics research community:

\begin{enumerate}
    \item \textbf{Novel Dataset Creation}: We introduce \textbf{ForensicsData}, a comprehensive synthetic dataset that accurately reflects malware behavior patterns across diverse Windows-based execution environments. The dataset incorporates rich Question-Context-Answer annotations specifically tailored for digital forensics applications, providing a valuable resource for training and evaluating forensic analysis tools.
    \item \textbf{Scalable LLM-Driven Annotation Pipeline}: We propose and implement an innovative, scalable pipeline that leverages multiple state-of-the-art LLMs to semantically annotate malware reports with structured forensic insights. This pipeline incorporates advanced prompt engineering techniques, parallel processing capabilities, and robust error handling mechanisms to ensure consistent, high-quality output generation across diverse malware families and attack vectors.
    \item \textbf{Comprehensive Validation Framework}: We develop and apply a multi-layered validation methodology that combines automated quality assessment techniques with expert evaluation protocols. This framework encompasses format validation, semantic deduplication, similarity filtering, and LLM-as-Judge evaluation to ensure the reliability, accuracy, and forensic relevance of generated datasets
\end{enumerate}

The remainder of this paper is organized as follows: Section II reviews related work in digital forensics, synthetic data generation, and LLM applications in cybersecurity. Section III details our data collection methodology and preprocessing procedures. Section IV presents our comprehensive methodology for dataset generation and validation. Section V discusses experimental results and performance comparisons across different LLM architectures, findings, limitations, and implications. Section VI concludes with a summary of contributions and outlines future research directions.

\section{Related Work}

Digital forensics (DF) refers to the collection, examination, and presentation of digital evidence. Technically and legally, the digital forensic landscape is becoming more complex due to the rapid proliferation of technologies like cloud services, embedded systems, and the Internet of Things (IoT) \cite{Yin_2025, Sharma_2023, Nelufule_2024}. Traditional forensic techniques, which were first created for more homogeneous and less voluminous data sources, are put to the test by the massive volumes of heterogeneous data generated by these contemporary computing environments \cite{Malik_2024, Narasimhan_2025}.

The lack of datasets appropriate for training and research is a major problem in digital forensics. Due to ethical, privacy, and legal restrictions, researchers are forced to use synthetic datasets instead of authentic forensic data. A good substitute that preserves privacy and permits scalable and repeatable experiments is the use of synthetic datasets \cite{Sharma_2025, Malik_2024, Zouhri2025WiFiQnA, Bellouch2025ParetoDQLMultiMDPICC, 10786584}. GPT and LLaMA, two recent developments in Large Language Models (LLMs), have shown promise in producing realistic and contextually rich synthetic forensic data \cite{Bai_2024, Xin_2024}. The practical implementation of these generative capabilities in digital forensic workflows is demonstrated by models such as \emph{ForensicLLM} \cite{Sharma_2025} and methods for producing comprehensive forensic reports \cite{Michelet_2023}. Furthermore, the synthesis of logs simulating cyber threats \cite{Chernyshev_2023} and the integration of LLM outputs with anomaly detection frameworks and explainable AI tools \cite{Ali_2023, Yin_2025} demonstrate promising avenues for automating threat triage, investigation, and analysis.

It is crucial to guarantee the stability and dependability of synthetic forensic data produced by LLMs. This calls for exacting standards and evaluation procedures. Initiatives like SciFaultyQA \cite{Kundu_2024}, CyberMetric \cite{Tihanyi_2024}, and CTIBench \cite{Alam_2024} offer structured datasets for assessing the fault detection, cybersecurity domain knowledge, and threat intelligence reasoning skills of LLM-generated outputs. Tools like LongCite \cite{Zhang_2024} and LongWriter \cite{Bai_2024}, which are expressly made to improve coherence in long-context text, are essential for enhancing the quality and accuracy of forensic reports that are produced, especially when it comes to citation accuracy and traceability. Maintaining the evidential rigor and admissibility necessary for forensic documentation requires the use of citation-aware frameworks, as demonstrated by the work of Gao et al. \cite{Gao_2023}.

Additionally, cross-domain developments offer important insights on enhancing LLM outputs in digital forensics. Advances in fields like AI safety \cite{Li_2024}, recommender systems \cite{Liang_2024}, and code synthesis \cite{Yun_2024} provide transferable approaches to guarantee safety, explainability, and controllability in LLM-driven synthetic data generation. Furthermore, domain-specific data augmentation methods investigated in intricate domains like chip design \cite{Chang_2024} demonstrate how LLMs may be tailored to certain technical contexts, indicating that DF may have a comparable potential.

Despite their considerable promise, LLMs and synthetic datasets face persistent challenges, including factual consistency, rigorous evaluation, and legal admissibility in digital forensics. Addressing these challenges requires sustained interdisciplinary research to align AI-driven tools with forensic standards, ensuring both technological innovation and adherence to legal frameworks.

While existing datasets have significantly contributed to digital forensic research, none specifically address malware behaviors in structured Question-Context-Answer (Q-C-A) format as comprehensively as \textbf{ForensicsData}. This gap underscores the importance and novelty of our contribution in facilitating structured forensic investigations and model training.

\section{Data Collection}
To create a comprehensive and up-to-date database of malware behavior, we sourced 1,500 execution reports from the publicly accessible \textit{ANY.RUN}\footnote{https://any.run/} malware analysis platform. It provides interactive sandbox environments for dynamic malware analysis and hosts over 10 million user-contributed malware and benign execution traces. The platform offers detailed logs of process execution, file system activity, network communications, and behavioral indicators associated with both malicious and benign software.
\subsection{Data Source}
The reports selected for this study were constrained to samples submitted in 2025 to ensure the relevance and currency of behavioral patterns. The dataset includes:
\begin{itemize}
    \item \textbf{15 Malware Families}: covering a diverse range of threats (e.g., Remote Access Trojans, credential stealers, ransomware, and loaders).
    \item \textbf{Benign Samples}: merged to support comparative analysis and multi-label classification tasks.

    \item \textbf{Uniform Distribution}: across families to minimize class imbalance and improve the generalizability in the dataset.
\end{itemize}

\begin{table}[H]
\centering
\caption{Malware Classes}
\begin{tabular}{ll}
\toprule
\textbf{Malware Type} & \textbf{Malware Family} \\
\midrule
Banker & qbot \quad trickbot \\
Benign & benign \\
Botnet & sality \\
Infostealer & formbook \quad hawkeye \\
Infostealer/Loader & amadey \\
Ransomware & gandcrab \quad wannacry \\
RAT & nanocore \quad xworm \quad remcos \\
Stealer & agenttesla \quad lumma \\
Trojan & emotet \\
\bottomrule
\end{tabular}
\end{table}

\begin{table}[H]
\centering
\caption{Distribution of Malware Family}
\begin{tabular}{l l l}
\toprule
\textbf{Malware Type} & \textbf{File Count} & \textbf{Percentage} \\
\midrule
agenttesla & 73 & 6.6\% \\
amadey & 40 & 3.6\% \\
emotet & 33 & 3.0\% \\
gandcrab & 100 & 9.1\% \\
lumma & 47 & 4.3\% \\
nanocore & 76 & 6.9\% \\
sality & 83 & 7.5\% \\
trickbot & 64 & 5.8\% \\
formbook & 86 & 7.8\% \\
hawkeye & 54 & 4.9\% \\
qbot & 69 & 6.3\% \\
remcos & 69 & 6.3\% \\
wannacry & 36 & 3.3\% \\
xworm & 72 & 6.5\% \\
benign & 150 & 13.6\% \\
\bottomrule
\end{tabular}
\end{table}

\subsection{Data Extraction and Preprocessing}
\begin{figure}[H]
    \centering
    \includegraphics[width=1\linewidth]{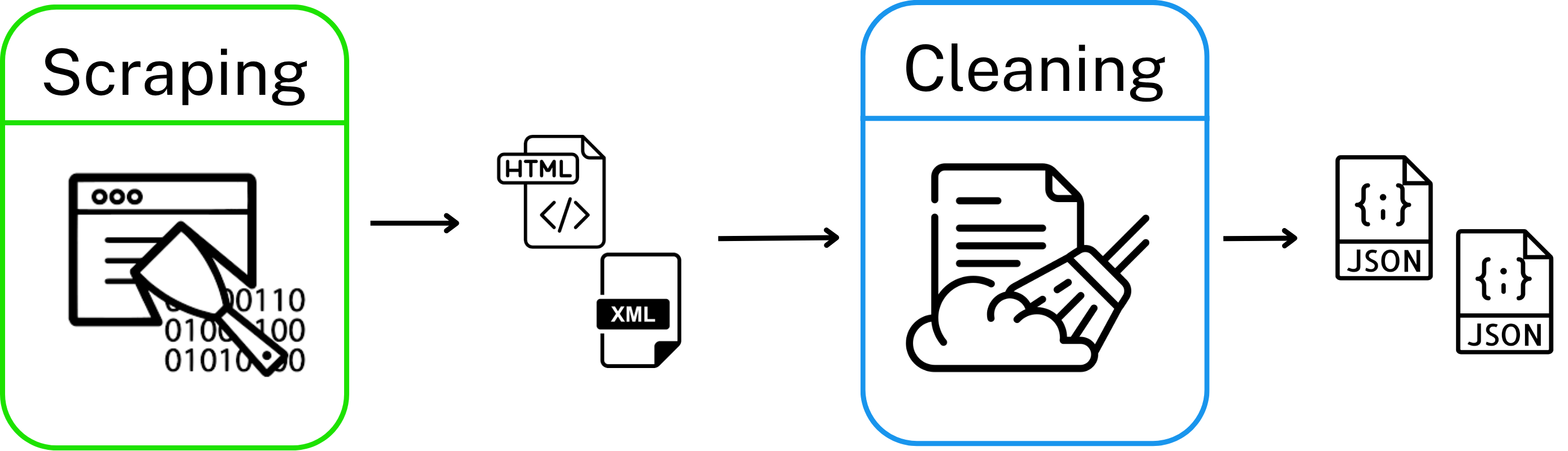}
    \caption{Data Scraping and Cleaning }
    \label{fig:cleaning_process}
\end{figure}

The original reports were available in HTML and semi-structured XML formats. To prepare them for LLM processing, we performed the following preprocessing steps:
\begin{itemize}
    \item\textbf{Parsing and Extraction:} Leveraging a custom pipeline built using \textit{BeautifulSoup} and \textit{lxml}, we extracted relevant behavioral sections from each report, including metadata (hashes, file names, verdicts, tags), indicators of compromise, behavior activities, behavior graphs, network indicators, and process trees.

    \item\textbf{Noise Removal:} Background processes unrelated to malware behavior were identified and removed. Files without meaningful content or containing incomplete information were also filtered out to reduce ambiguity.

    \item\textbf{Standardization}: All reports were converted to a consistent \texttt{JSON format} with standardized filed names and value formats to facilitate automated processing.

    \item\textbf{De-duplication}: The dataset was checked for repeated samples, and removed duplication malware report based on SHA-256 hash comparisons.

The preprocessing phase resulted in clean, structured dataset in JSON format (see Figure~\ref{fig:cleaning_process}) ready for Question and Answer generation by LLMs
\end{itemize}
\section{Methodology }
 
The core objective of this work is to transform raw malware execution reports into a high-quality synthetic dataset of question-context-answer (Q-C-A) triples for digital forensics research. To achieve this, we designed a multi-phase methodology comprising structured data transformation, semantic annotation via large language models (LLMs), and rigorous validation of outputs. The methodology is organized into two main components: 

\subsection{The LLM Annotation Pipeline and Validation Techniques.}

\begin{figure*}[t]
    \centering
    \includegraphics[width=0.75\linewidth]{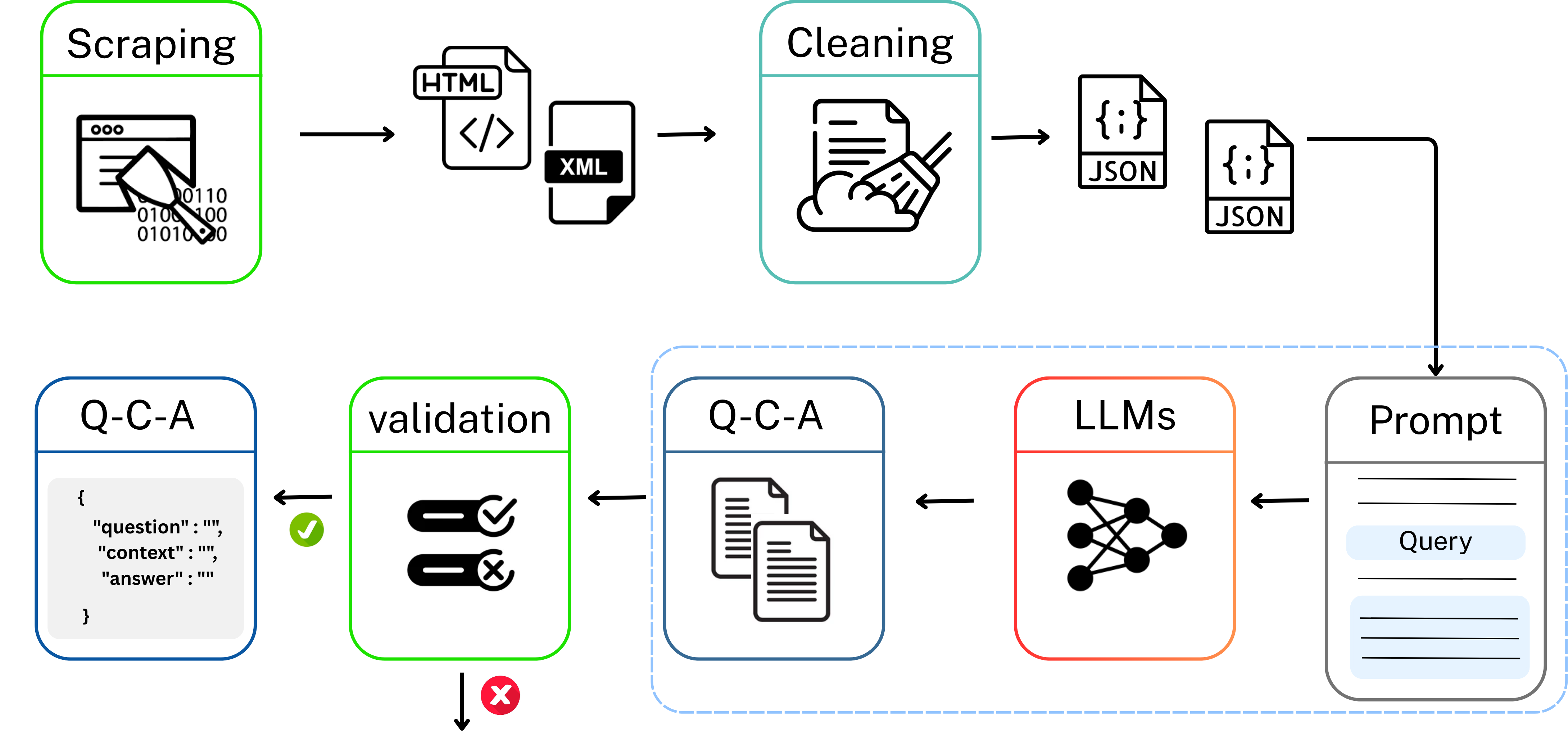}
    \caption{pipeline for generating structured datasets Question-Context-Answer (Q-C-A) triples}
    \label{fig:enter-label}
\end{figure*}


\subsubsection{LLMs Pipeline: }
\begin{figure}[H]
    \centering
    \includegraphics[width=1\linewidth]{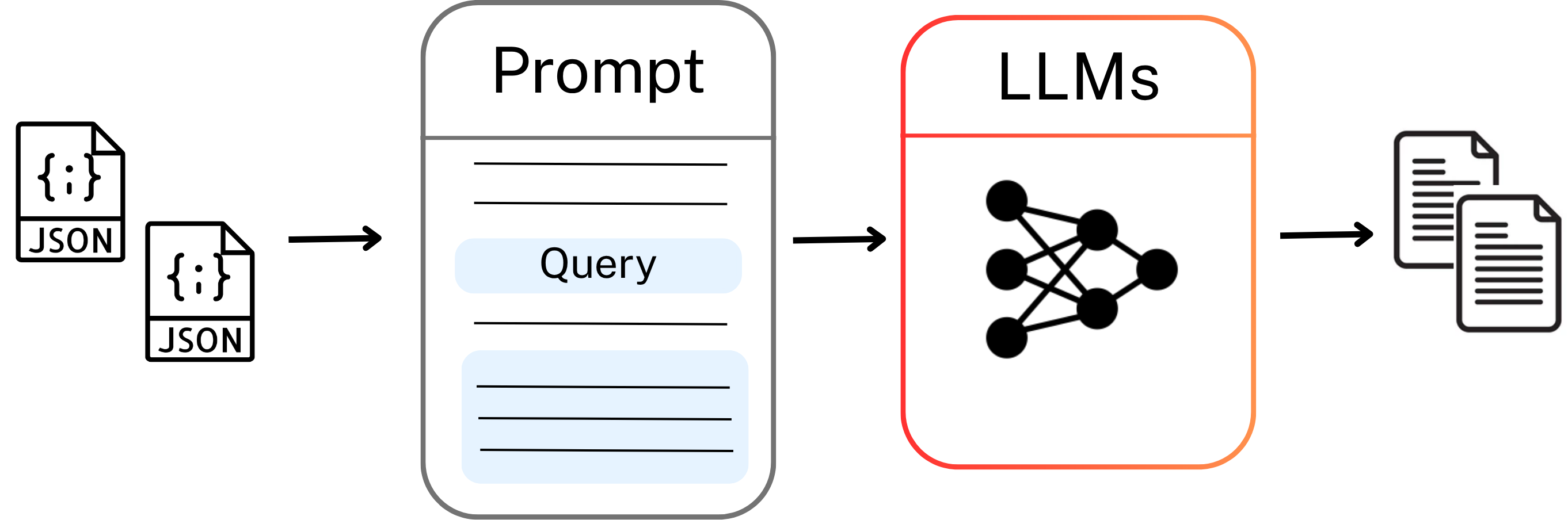}
    \caption{Create Question Context Answer by a LLM}
    \label{fig:generateqca}
\end{figure}
The \textbf{ForensicsData} generation process involves transforming structured JSON representations of malware reports into Question-Context-Answer (Q-C-A) triples through a scalable and parallelized LLM annotation pipeline. The pipeline consists of the following stages (see Figure~\ref{fig:generateqca}):

\subsubsection{Prompt Engineering: }
Each JSON report was passed to an LLM along with a prompt instructing it to generate contextually grounded Question-Context-Answer (Q-C-A) triples. Prompts were designed with care to return diverse forensic insights essential for \textbf{ForensicsData} across five dimensions.

\begin{itemize}
    \item Malware identification and metadata.
    \item Technical indicators of compromise.
    \item Suspicious Behavioral patterns and techniques.
    \item Malicious Behavioral patterns and techniques.
    \item Impact assessment and mitigation strategies.
\end{itemize}
    
\subsubsection{LLM Selection and Allocation: }

To capture diversity in generation style and reasoning patterns, five different LLMs were used:

\begin{table}[H]
\centering
\caption{Comparison of Language Models}
\begin{tabular}{l l l l}
\toprule
\textbf{Model } &  \textbf{Parameters} & \textbf{Context } \\
\midrule
Mistral 8B  & $\approx$ 7.3 billion & 32k tokens \\
LLaMA 3–70B &  $\approx$ 70 billion & 8k tokens \\
DeepSeek V3  & 33 billion & 128k tokens \\
Qwen-QWQ-32B & $\approx$ 32 billion & 32k tokens \\
Gemini 2.0 Flash &  Not disclosed & 1M tokens \\
\bottomrule
\end{tabular}
\end{table}

Each model processed 20\% of the dataset (approximately 300 reports) and generated five Q-C-A triples per report. This stratified approach ensured balanced representation across models and malware families.

\textbf{Output Processing}: The generated Q-C-A triples were automatically aggregated and formatted according to the predefined JSON schema, resulting in approximately
5,000 initial Q-C-A triples (1000-1200 Q-C-A triples per model). These triples were compiled into the \textbf{ForensicsData} dataset.
The pipeline incorporated error handling and retry mechanisms to address generation failures or unexpected output

\subsection{Multi-layered Validation for Synthetic Datasets} 
Given the synthetic nature of the dataset, validating the quality, coherence, and utility of generated Q-C-A triples was essential. We employed a multi-layered validation strategy:
\begin{figure}[H]
    \centering
    \includegraphics[width=1\linewidth]{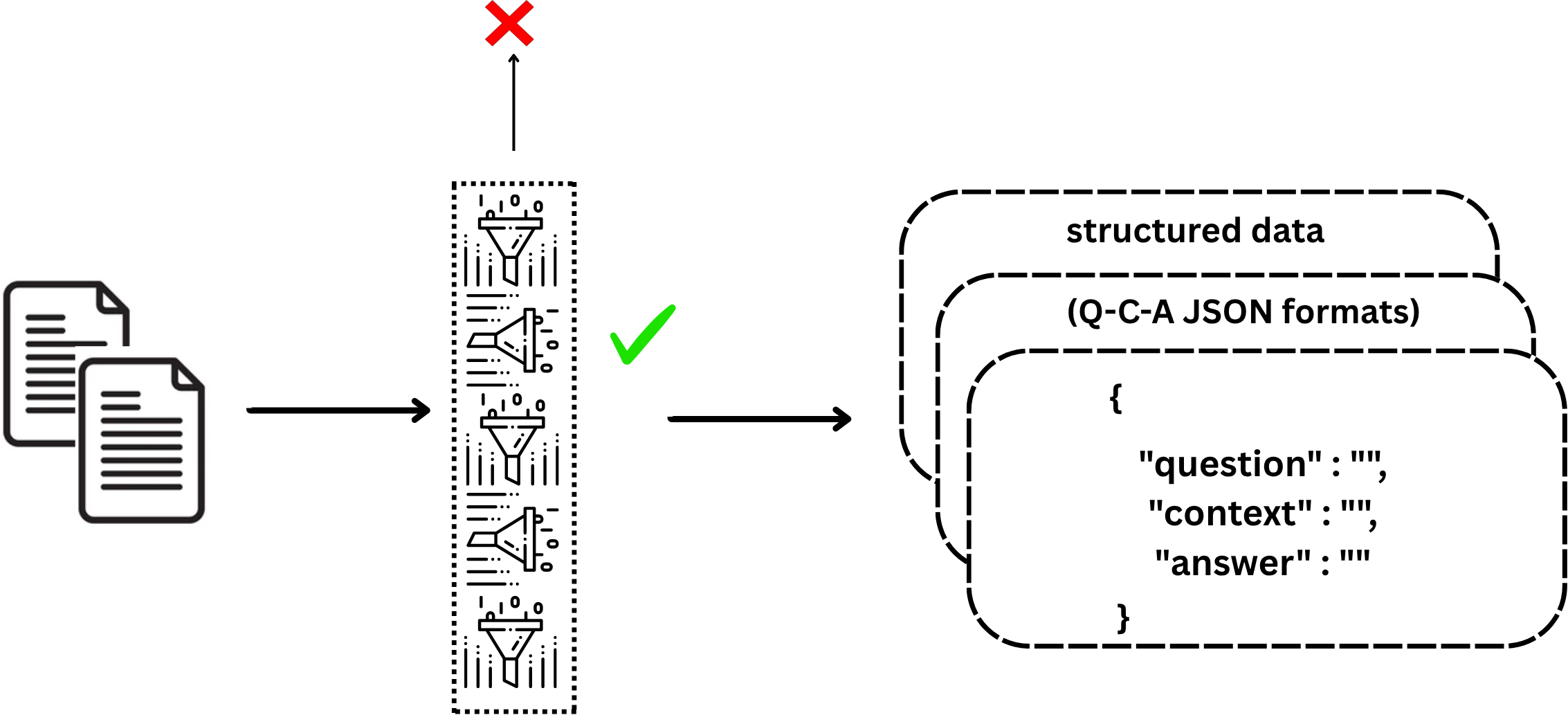}
    \caption{Framework of Validation Dataset Generation}
    \label{fig:validation_1}
\end{figure}
\textbf{1. Format Validation via Pydantic library}

Each generated Q-C-A entry was parsed using the Pydantic library to enforce schema correctness, type safety, and structural consistency across the dataset. The schema enforced the following structure:

\begin{verbatim}
{
  "question": "string",
  "context": "string",
  "answer": "string"
}
\end{verbatim}

\textbf{2. Deduplication and Similarity Filtering}

We employed vector-based similarity analysis using the \texttt{all-MiniLM-L6-v2} Embedding Model from the \texttt{sentence-transformers} library to generate dense sentence embeddings. Cosine similarity scores, computed via scikit-learn, were used to identify questions with high semantic overlap. Questions with a similarity score above 0.9 were flagged and removed to eliminate redundancy.
Given a Question \(Q \), the \texttt{all-MiniLM-L6-v2} model produces an embedding vector \(\mathbf{e}_q \in \mathbb{R}^d \), where \(d = 385 \) is the embedding dimension.

Cosine similarity between two embedding vectors \(\mathbf{e}_a \) and \(\mathbf{e}_b \) is computed as:

\begin{equation}
\text{cosine\_similarity}(\mathbf{e}_a, \mathbf{e}_b) = 
\frac{\mathbf{e}_a \cdot \mathbf{e}_b}{\|\mathbf{e}_a\| \|\mathbf{e}_b\|}
\label{eq:cosine}
\end{equation}

where, $\cdot$ denotes the dot product (inner product) of the two vectors, and $\|\mathbf{e}_a\|$ and $\|\mathbf{e}_b\|$ are their Euclidean norms.
\begin{figure}[H]
    \centering
    \includegraphics[width=1\linewidth]{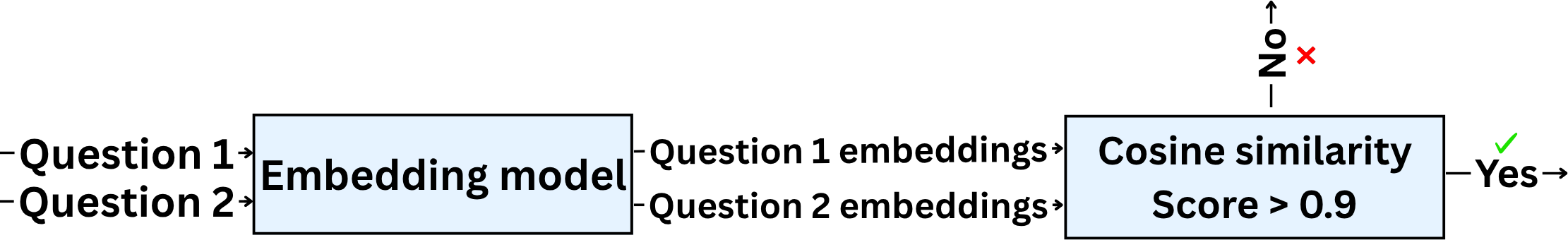}
    \caption{Workflow for Semantic Questions Filtering Using Embedding Model}
    \label{fig:enter-label}
\end{figure}

\textbf{3. LLM-as-Judge Evaluation}
A high-performing language model (\textbf{Gemini Advanced 2.0 flash}) was used as an automated evaluator to assess the quality of question-context-answer triples. The evaluation followed a structured prompt that guided the LLM to analyze each Q-C-A triples across four detailed criteria:

\begin{itemize}
    \item \textbf{Logical Validity and Realism} Determines whether the question is logically sound and based on realistic technical or security scenarios.

    \item \textbf{Relevance} Assesses if the answer directly addresses the question in a meaningful and appropriate way.

    \item \textbf{Completeness} 
    Evaluates whether the answer fully covers all aspects of the question.

    \item \textbf{Consistency and Absence of Hallucinations} 
    Verifies that the answer is internally consistent and free from hallucinated (fabricated) information.

\end{itemize}
For each criterion, the model provided a clear verdict (Yes/No, 1/0) along with a brief justification. Q-C-A triples that failed to meet one or more of the criteria were either revised or excluded from the dataset to maintain quality and reliability.

\section{Results and Discussion }
\begin{table*}[htbp]
    \centering
    \caption{Performance Benchmarking of LLMs Used for Generating \textbf{ForensicsData}}
    \resizebox{\textwidth}{!}{%
    \begin{tabular}{lcccccc}
    \toprule
    \textbf{Model} & \textbf{Provider} & \textbf{Parameters} & \textbf{Context Windows} & \textbf{Latency (s)} & \textbf{Cost Input/Output (\$/1K tokens)} & \textbf{JSON Validity (\%)} \\
    \midrule
    Qwen-QWQ-32B     & GroqCloud   & 32B        & 32k  & 250  & 0.00029 / 0.00039 & 99.3 \\
    LLaMA 3–70B      & GroqCloud   & 70B        & 8k   & 364  & 0.00059 / 0.00079 & 98.6 \\
    Gemini 2.0 Flash & Google      & Undisclosed& 1M   & 425  & 0.0001 / 0.0004   & 99.8 \\
    Mistral 8B       & OpenRouter  & 7.3B       & 32k  & 750  & 0.0001 / 0.0001   & 97.2 \\
    DeepSeek V3      & OpenRouter  & 33B        & 128k & 1667 & 0.00038 / 0.00089 & 99.5 \\
    \bottomrule
    \end{tabular}%
    }
\end{table*}

Here, we compare the performance of the five large language models used to produce \textbf{ForensicsData} dataset from formal JSON reports. Each model was responsible for enriching approximately 20\% of the dataset semantically by producing five Q-C-A triples per report. We assess the usability of this generation process in the real world.

\subsection{\textbf{Performance Benchmarking of LLMs}}
To establish the pipeline operation efficiency of the pipeline in terms of generating datasets, we compared all the five LLMs on four aspects:

\subsubsection{Latency}
Comparing with models on host traditional infrastructures, GroqCloud-supported larger models such as the Qwen QWQ 32B and LLaMA 3-70B exhibit considerably lower latency and higher throughput. GroqCloud's use of application-specific hardware (GroqChips) designed for ultra-low latency inference is the performance advantage here. In contrast, vendors such as Mistral and Gemini rely on NVIDIA GPUs (such as the A100) and TPUs (such as the TPUv4), respectively, which are not as latency-optimized as Groq's own hardware.
 
\subsubsection{Model Throuphput Performance}
Qwen QWQ-32B and LLama3 70B achieved the highest throughput, supported by the optimized inferenced infrastructure provided by GroqCLoud.

\subsubsection{Models Cost}
Maximum output at cost-effective price is provided by the Qwen-QWQ-32B with the best trade-off between API cost and performance of output. The most cost-efficient models, Mistral 8B and Gemini Flash, are suited for usage scenarios where money is tight. In contrast, though LLaMA 3–70B excels in performance, it has the highest cost of output and is therefore not ideally suited for applications where cost-effectiveness takes priority.
 
\subsubsection{Output format validity}
Over 97\% of generated outputs from every model were successfully validated by Pydantic schema, illustrating the success of structural formatting and prompt engineering. Compliance with instructions, however, was inconsistent: while Mistral 8B produced syntactically correct output, it occasionally failed to cover all semantic meaning of prompts, showing weaker instruction-following skills in spite of its structural correctness.

\subsection{\textbf{Synthetic Dataset Quality Control and Validation}}
In order to assure the usability and integrity of the resulting synthetic \textbf{ForensicsData} dataset produced by the five large language models, we performed a series of quality control procedures ensuring format validity, semantic variety, logical consistency, and redundancy. The following is a summary of the main dataset characteristics resulting from these post-generation tests.

\subsubsection{Format Validation via Pydantic library}
 98.68\% of the entries passed schema validation without requiring any modification. The remaining entries were either auto-corrected or discarded if invalid.

 \subsubsection{Deduplication and Similarity Filtering}

Deduplication and similarity filtering with the MiniLM-L6-V2 model at 0.90 cosine similarity threshold ensured that no Q-C-A triples were marked as duplicates. This indicates that the LLMs generated semantically different, context-sensitively aware content consistently even in similar malware samples, indicating high semantic diversity and ability to generate different outputs even in highly similar input scenarios.

\subsubsection{Logical Coherence Check via LLM-as-Judge}

\begin{table}[H]
\centering 
\begin{tabular}{@{}lcc@{}}
\toprule
\textbf{Criterion} & \textbf{Q-C-A Passing} & \textbf{Validation Rate} \\
\midrule
Logical Validity     & 5,000 / 5,000 & 100.0\% \\
Relevance            & 5,000 / 5,000 & 100.0\% \\
Completeness         & 4850 / 5,000 & 98.0\%  \\
Hallucination-Free   & 5,000 / 5,000 & 100.0\% \\
\bottomrule
\end{tabular}

\caption{Validation results of Q-C-A generation}
\label{tab:validation_results}
\end{table} 

These findings confirm that the vast majority of Q-C-A triples are logically correct, accurate, and contextually applicable. The sole meaningful constraint was completeness—around 2\% of answers left questions partially answered, typically glossing over technical details like file paths, process IDs, or behavior descriptions. Gemini 2.0 Flash, a validator based on an LLM, verified 100\% logical correctness and hallucination-free generation and also detected that 2\% of triples were completely complete. This shortfall offers potential for improvement through more directed prompting or post-generation revision.

\subsection{Strengths and Practical Implications}
The methodology and resulting dataset present several notable strengths:

\begin{enumerate}
    \item \textbf{Realistic Behavioral Coverage}: Reports from 15 malware families ensure coverage across multiple malware categories, including RATs, Banker, Trajon, stealers, botnet, and ransomware.
    
    \item \textbf{Recency and Relevance}: All source reports were collected in 2025, aligning the dataset with modern malware techniques and infrastructure (e.g., C2 communication patterns, evasion tactics).
    
    \item \textbf{Structured and Validated Outputs}: JSON-based Q-C-A triples were schema-validated using Pydantic, and semantic coherence was verified using an independent LLM judge (Gemini 2.0 flash).
    
    \item \textbf{Cross-Model Generation}: Using five LLMs improved stylistic and conceptual variety, enhancing the dataset’s richness and generalizability.
    
    \item \textbf{Ethical Safety}: By generating descriptive Q-C-A triples rather than executable code or raw malware, the pipeline maintains ethical integrity while enabling security research.
\end{enumerate}

\subsection{Limitations and Challenges}
Despite its strengths, the approach has certain limitations:

\begin{enumerate}
    \item \textbf{Instruction Sensitivity}: Some models, particularly Mistral 8B, adhered poorly to detailed instructions, requiring additional filtering or re-generation in post-processing.
    
    \item \textbf{Source Dependency}: Output quality is inherently limited by the completeness and clarity of the original ANY.RUN report content.
\end{enumerate}

\subsection{Hallucination and Bias Considerations}
LLMs are known to introduce hallucinations in generated outputs. We addressed these issues explicitly.

\subsubsection*{Hallucination Mitigation}
While Gemini 2.0 Flash found no hallucinations in Q-C-A triples, we proactively implemented the following safeguards:

\begin{itemize}
    \item \textbf{Prompt constraints}: LLMs were instructed to only use content available in the report JSON.
    \item \textbf{Schema enforcement}: Ensured structured and bounded outputs.
    \item \textbf{LLM-as-Judge filtering}: Discarded or revised outputs judged to contain inconsistencies.
\end{itemize}

\section{Conclusion and Perspectives}
The article describes a scalable approach to creating a synthetic malware behavior \textbf{ForensicsData} dataset for digital forensics that addresses dataset scarcity while maintaining privacy compliance. Using 1,500 malware and benign reports from ANY.RUN, we used five big language models to generate 5,000 question-context-answer triples covering malware identification, technical indicators, behavioral patterns, and mitigatienter-labelon strategies.
A multi-layered validation method, comprising format checks, deduplication, similarity filtering, and LLM-as-Judge review, revealed high dataset quality, with 100\% logical validity and relevance. However, 2\% of responses indicated minor incompleteness. Performance benchmarking revealed trade-offs amongst models in terms of cost, speed, and instruction adherence. The generated dataset and pipeline provide the foundation for intelligent forensic tools and reproducible operations. Future work will include fine-tuning LLMs for particular forensic tasks, creating Retrieval-Augmented Generation systems, investigating multi-agent analysis, and expanding the dataset to encompass more platforms and sample sizes.

\bibliographystyle{IEEEtran} 

\begin{thebibliography}{10}
\providecommand{\url}[1]{#1}
\csname url@samestyle\endcsname
\providecommand{\newblock}{\relax}
\providecommand{\bibinfo}[2]{#2}
\providecommand{\BIBentrySTDinterwordspacing}{\spaceskip=0pt\relax}
\providecommand{\BIBentryALTinterwordstretchfactor}{4}
\providecommand{\BIBentryALTinterwordspacing}{\spaceskip=\fontdimen2\font plus
\BIBentryALTinterwordstretchfactor\fontdimen3\font minus \fontdimen4\font\relax}
\providecommand{\BIBforeignlanguage}[2]{{%
\expandafter\ifx\csname l@#1\endcsname\relax
\typeout{** WARNING: IEEEtran.bst: No hyphenation pattern has been}%
\typeout{** loaded for the language `#1'. Using the pattern for}%
\typeout{** the default language instead.}%
\else
\language=\csname l@#1\endcsname
\fi
#2}}
\providecommand{\BIBdecl}{\relax}
\BIBdecl

\bibitem{Yin_2025}
Z.~Yin, Z.~Wang, W.~Xu, J.~Zhuang, P.~Mozumder, A.~Smith, and W.~Zhang, ``Digital forensics in the age of large language models,'' \emph{null}, 2025.

\bibitem{Sharma_2023}
P.~Sharma and L.~Awasthi, ``Next-generation digital forensics challenges and evidence preservation framework for iot devices,'' \emph{International Journal of Next-Generation Computing}, 2023.

\bibitem{Nelufule_2024}
N.~Nelufule, T.~Singano, and M.~Masango, ``A comprehensive exploration of digital forensics investigations in embedded systems, ubiquitous computing, fog computing, and edge computing,'' \emph{2024 International Conference on Artificial Intelligence, Big Data, Computing and Data Communication Systems (icABCD)}, 2024.

\bibitem{Malik_2024}
A.~W. Malik, D.~S. Bhatti, T.-J. Park, H.~U. Ishtiaq, J.-C. Ryou, and K.-I. Kim, ``Cloud digital forensics: Beyond tools, techniques, and challenges,'' \emph{Italian National Conference on Sensors}, 2024.

\bibitem{Narasimhan_2025}
P.~Narasimhan and D.~N. Kala, ``Emerging trends in digital forensics : Investigating cybercrime,'' \emph{International Journal of Scientific Research in Computer Science Engineering and Information Technology}, 2025.

\bibitem{Sharma_2025}
B.~Sharma, J.~Ghawaly, K.~McCleary, A.~Webb, and I.~M. Baggili, ``Forensicllm: A local large language model for digital forensics,'' \emph{Digital Investigation. The International Journal of Digital Forensics and Incident Response}, 2025.

\bibitem{Zouhri2025WiFiQnA}
A.~Zouhri, L.~Zitoune, and I.~Lahsen-Cherif, ``{WiFiQnA}: a {WiFi} dataset for large language models,'' \url{https://hal.science/hal-05146992}, 2025, hAL Id: hal-05146992.

\bibitem{Bellouch2025ParetoDQLMultiMDPICC}
\BIBentryALTinterwordspacing
M.~Bellouch, L.~Zitoune, I.~Lahsen-Cherif, and V.~V{\`e}que, ``Pareto {DQL-MultiMDP} sub-controllers for load balancing in large and dynamic {WiFi} networks,'' in \emph{IEEE International Conference on Communications (ICC)}, Montr{\'e}al, Canada, Jun. 2025, hAL Id: hal-04952832. [Online]. Available: \url{https://hal.science/hal-04952832}
\BIBentrySTDinterwordspacing

\bibitem{10786584}
M.~Bellouch, L.~Zitoune, I.~Lahsen-Cherif, and V.~Vèque, ``Load balancing in large wifi networks using dql-multimdp with constrained clustering,'' in \emph{2024 32nd International Conference on Modeling, Analysis and Simulation of Computer and Telecommunication Systems (MASCOTS)}, 2024, pp. 1--8.

\bibitem{Bai_2024}
Y.~Bai, J.~Zhang, X.~Lv, L.~Zheng, S.~Zhu, L.~Hou, Y.~Dong, J.~Tang, and J.~Li, ``Longwriter: Unleashing 10,000+ word generation from long context llms,'' \emph{arXiv.org}, 2024.

\bibitem{Xin_2024}
H.~Xin, D.~Guo, Z.~Shao, Z.~Ren, Q.~Zhu, B.~L.~B. Liu), C.~Ruan, W.~Li, and X.~Liang, ``Deepseek-prover: Advancing theorem proving in llms through large-scale synthetic data,'' \emph{arXiv.org}, 2024.

\bibitem{Michelet_2023}
G.~Michelet and F.~Breitinger, ``Chatgpt, llama, can you write my report? an experiment on assisted digital forensics reports written using (local) large language models,'' \emph{Forensic Science International: Digital Investigation}, 2023.

\bibitem{Chernyshev_2023}
M.~Chernyshev, Z.~A. Baig, and R.~Doss, ``Towards large language model (llm) forensics using llm-based invocation log analysis,'' \emph{LAMPS@CCS}, 2023.

\bibitem{Ali_2023}
T.~M. Ali and P.~Kostakos, ``Huntgpt: Integrating machine learning-based anomaly detection and explainable ai with large language models (llms),'' \emph{arXiv.org}, 2023.

\bibitem{Kundu_2024}
D.~Kundu, ``Scifaultyqa: Benchmarking llms on faulty science question detection with a gan-inspired approach to synthetic dataset generation,'' \emph{arXiv.org}, 2024.

\bibitem{Tihanyi_2024}
N.~Tihanyi, M.~Ferrag, R.~Jain, T.~Bisztray, and M.~Debbah, ``Cybermetric: A benchmark dataset based on retrieval-augmented generation for evaluating llms in cybersecurity knowledge,'' \emph{Computer Science Symposium in Russia}, 2024.

\bibitem{Alam_2024}
M.~T. Alam, D.~Bhusal, L.~Nguyen, and N.~Rastogi, ``Ctibench: A benchmark for evaluating llms in cyber threat intelligence,'' \emph{Neural Information Processing Systems}, 2024.

\bibitem{Zhang_2024}
J.~Zhang, Y.~Bai, X.~Lv, W.~Gu, D.~Liu, M.~Zou, S.~Cao, L.~Hou, Y.~Dong, L.~Feng, and J.~Li, ``Longcite: Enabling llms to generate fine-grained citations in long-context qa,'' \emph{arXiv.org}, 2024.

\bibitem{Gao_2023}
T.~Gao, H.~Yen, J.~Yu, and D.~Chen, ``Enabling large language models to generate text with citations,'' \emph{Conference on Empirical Methods in Natural Language Processing}, 2023.

\bibitem{Li_2024}
M.~Li, J.~Chen, L.~Chen, and T.~Zhou, ``Can llms speak for diverse people? tuning llms via debate to generate controllable controversial statements,'' \emph{Annual Meeting of the Association for Computational Linguistics}, 2024.

\bibitem{Liang_2024}
T.~Liang, C.~Jin, L.~Wang, W.~Fan, C.~Xia, K.~Chen, and Y.~Yin, ``Llm-redial: A large-scale dataset for conversational recommender systems created from user behaviors with llms,'' \emph{Annual Meeting of the Association for Computational Linguistics}, 2024.

\bibitem{Yun_2024}
S.~Yun, H.~Lin, R.~Thushara, M.~Q. Bhat, Y.~Wang, Z.~Jiang, M.~Deng, J.~Wang, T.~Tao, J.~Li, H.~Li, P.~Nakov, T.~Baldwin, Z.~Liu, E.~P. Xing, X.~Liang, and Z.~Shen, ``Web2code: A large-scale webpage-to-code dataset and evaluation framework for multimodal llms,'' \emph{Neural Information Processing Systems}, 2024.

\bibitem{Chang_2024}
K.~Chang, K.~Wang, N.~Yang, Y.~Wang, D.~Jin, W.~Zhu, Z.~Chen, C.~Li, H.~Yan, Y.~Zhou, Z.~Zhao, Y.~Cheng, Y.~Pan, Y.~Liu, M.~Wang, S.~Liang, Y.~Han, H.~Li, and X.~Li, ``Data is all you need: Finetuning llms for chip design via an automated design-data augmentation framework,'' \emph{Design Automation Conference}, 2024.

\end{thebibliography}

\end{document}